\documentclass[sigconf,screen]{acmart}

\setcopyright{none}
\renewcommand\footnotetextcopyrightpermission[1]{} 

\usepackage{geometry}
\usepackage{algorithmic}
\usepackage{graphicx}
\usepackage{textcomp}
\usepackage{xcolor}
\usepackage{stfloats}
\usepackage{url}
\usepackage{enumitem}
\usepackage{booktabs} 
\usepackage{multirow} 
\usepackage{color, colortbl}
\definecolor{Gray}{gray}{0.9}

\usepackage{color,soul}

\usepackage{pifont}

\usepackage{tikz} 

\usepackage{tcolorbox}

\tikzstyle{every number}=[draw=white] 

\usepackage[ruled,vlined,noend]{algorithm2e} 
\usepackage[ruled,vlined,noend]{algorithm2e}

\copyrightyear{2023} 
\acmYear{2023} 
\setcopyright{acmlicensed}\acmConference[ARES 2023]{The 18th International Conference on Availability, Reliability and Security}{August 29-September 1, 2023}{Benevento, Italy}
\acmBooktitle{The 18th International Conference on Availability, Reliability and Security (ARES 2023), August 29-September 1, 2023, Benevento, Italy}
\acmPrice{15.00}
\acmDOI{10.1145/3600160.3604985}
\acmISBN{979-8-4007-0772-8/23/08}

\begin{document}

    



\author{Alpesh Bhudia} 
\affiliation{
\institution{\normalsize Royal Holloway, University of London}
\city{Egham}
\country{United Kingdom \\ alpesh.bhudia.2018@live.rhul.ac.uk}
}

\author{Anna Cartwright}
\affiliation{\normalsize \institution{Oxford Brookes University}
\city{Oxford}
\country{United Kingdom \\ a.cartwright@brookes.ac.uk}
}

\author{Edward Cartwright}
\affiliation{\normalsize 
\institution{De Montfort University}
\city{Leicester}
\country{United Kingdom \\ edward.cartwright@dmu.ac.uk}
}

\author{Darren Hurley-Smith}
\affiliation{\normalsize 
\institution{Royal Holloway, University of London}
\city{Egham}
\country{United Kingdom \\ darren.hurley-smith@rhul.ac.uk}
}
\author{Julio Hernandez-Castro}
\affiliation{\normalsize 
\institution{University of Kent}
\city{Canterbury}
\country{United Kingdom \\ j.c.hernandez-castro@kent.ac.uk}
}

\title{Game Theoretic Modelling of a Ransom and Extortion \\ Attack on Ethereum Validators}

\begin{abstract}
Consensus algorithms facilitate agreement on and resolution of blockchain functions, such as smart contracts and transactions.  Ethereum uses a Proof-of-Stake (PoS) consensus mechanism, which depends on financial incentives to ensure that \emph{validators} perform certain duties and do not act maliciously. Should a validator attempt to defraud the system, legitimate validators will identify this and then staked cryptocurrency is `burned' through a process of \emph{slashing}.

In this paper, we show that an attacker who has compromised a set of \emph{validators} could threaten to perform malicious actions that would result in slashing and thus, hold those validators to ransom. We use game theory to study how an attacker can coerce payment from a victim, for example by deploying a smart contract to provide a root of trust shared between attacker and victim during the extortion process. Our game theoretic model finds that it is in the interests of the validators to fully pay the \emph{ransom} due to a lack of systemic protections for validators. Financial risk is solely placed on the victim during such an attack, with no mitigations available to them aside from capitulation (payment of ransom) in many scenarios. Such attacks could be disruptive to Ethereum and, likely, to many other PoS networks, if public trust in the validator system is eroded. We also discuss and evaluate potential mitigation measures arising from our analysis of the game theoretic model.


\end{abstract}


\maketitle
\pagestyle{plain}
\pagenumbering{gobble}


\section{Introduction}
Ethereum 2.0 differentiates itself from previous versions of Ethereum and most other cryptocurrencies, including Bitcoin, by departing from the Proof-of-Work (PoW) consensus mechanism. Proof-of-Stake (PoS) has become the de facto mode of Ethereum (eth), and so the 2.0 label has been retired and Ethereum now refers to the current PoS version of the protocol \cite{TheMerge82:online}. It uses a non-delegated proof-of-stake scheme to confirm transactions and add blocks to the chain. Proof-of-Stake requires that \emph{validators} are selected to agree on valid blocks of transactions, with a probability that is based on the amount they have staked in the network \cite{Proofofs94:online}. Validators are \emph{incentivized} to perform useful tasks by appropriate rewards and punishments.\footnote{Cardano technically implemented full proof-of-stake and full decentralisation of PoS first, however, it is significantly different to Ethereum's implementation, and there are some concerns about the integrity of the Delegated proof of stake \cite{gavzi2019}.} An advantage of PoS systems is reduced energy demands, as complex calculations (i.e. the cryptographic proof-of-work) are reduced, with the additional benefit of avoiding duplication of effort~\cite{Vranken2017}. Proof-of-stake (or a hybrid approach) can significantly reduce these energy costs and also potentially increase reliability and security by ruling out the ability for attackers to reduce their `cost of work', by stipulating a specific price of participation~\cite{Kiayias2017,Nguyen2019,Zhang2020}.

Crucial to any blockchain is resilience to manipulation and fraud. To quote from the Ethereum Foundation blog, `eth2 assumes validators will be lazy, take bribes, and that they will try to attack the system unless they are otherwise incentivised not to. Furthermore, the network is assumed not to be entirely reliable and that catastrophic events could force large numbers of validators to go offline' \cite{Proofofs923:online}. Where information is not reliant on centralised infrastructure, it is possible to add resilience through mass duplication~\cite{Beekhuizen2019}. It is, however, extraordinarily difficult for any system design process to predict and pre-empt all the possible attack scenarios, especially those defrauding consensus protocols. Some studies have, thus, explored a small number of vulnerabilities and attack vectors \cite{Chen2020,Deirmentzoglou2019,Gazi2018,Neuder2021,schwarz2021three}.

This work evaluates the impact of extortion attacks against the Ethereum validators. We use game theoretic modelling to determine the key decisions which shape the likelihood of payment of the ransom by the victim (validator owner). The purpose of this is to identify the specific systems and support mechanisms of Ethereum and the pressures that the attacker can apply. This allows us to map the decision flows between the attacker, victim, and Ethereum blockchain in the proposed extortion attack scenarios. Finally, this approach allows us to identify the contributing factors towards victims deciding to pay the ransom and the point in the decision making process at which such a decision becomes likely. This helps in identifying potential mitigations that could be implemented in future versions of Ethereum, with the aim of minimising damages to validators and the likelihood of ransom payment. This work has been undertaken as an independent study of Ethereum validator security, working closely with the Ethereum Foundation, who have been responsibly informed of all our findings.

\subsection{Background}
Ethereum (eth) validators each have two keys: a \emph{signing key} to perform a variety of duties, and a \emph{withdrawal key} to access the resource staked \cite{antonopoulos2018mastering, Kiayias2017}. An attacker who compromises both the signing and withdrawal key of a validator (via either known or zero-day vulnerabilities, e.g. \cite{ApacheLo94:online,CVE2020023:online}) can directly access and `steal' the validator's stake. This, however, can be avoided relatively easily if the validator keeps the withdrawal key fully secure, particularly since it only needs to be used for withdrawal and therefore isn't required (or advised) to be stored locally on the node. The signing key, by contrast, must be in constant use online and so is inevitably more vulnerable as it is stored locally on the validator. We focus here, therefore, on the more plausible scenario in which the attacker has compromised the signing key but not the withdrawal key of the validator. 

A criminal who has compromised the signing key could attempt a long range attack aimed at directly corrupting the blockchain consensus mechanism \cite{Deirmentzoglou2019}. Eth, and other crypto-currencies, are, however, actively designed to thwart this kind of attack. Such an attack is highly unlikely to succeed as it requires achieving control of an infeasibly large number (>50\% total population) of validators (by exploitation of vulnerable validators or expenditure of Eth to create attacker-controlled validators). 
As we shall show, extortion is a far more viable route to financial gain. Slashing is a process whereby a validator, who has seemingly acted maliciously against the eth network, is punished and incurs a significant loss of their committed resources, of at least 1/32 of the stake~\cite{unknown,fanti2019economics}. An attacker who has access to the signing key of a validator can force slashing and, therefore, can \emph{threaten} to force slashing unless a ransom is paid. In this paper, we demonstrate that a validator who faces losing money through slashing may, under certain conditions, rationally choose to pay the ransom. The validator can, thus, be extorted.

In formally evaluating the possibility of a ransom attack we need to consider incentive compatibility, in a game theoretic sense, for both the attacker (i.e. the criminal or insider acting maliciously) and the victim (i.e. the staker~\footnote{Throughout this paper, we use staker to define a 'unique eth wallet public key' rather than the individual behind the wallet, who may own many wallets, each of which may express different staking strategies.} whose validator(s) have been compromised)~\cite{Cartwright2019b}. Specifically, it will only be in the interest of the victim to pay a ransom if doing so will result (with sufficiently high probability) in a reduced financial loss and/or the return of control over the signing keys. The attackers, however, have incentives to renege on a deal by extracting ransom but then not returning control of the keys, allowing them to make a second or successive ransom demand. This incentive to renege means that, in order to adequately incentivize ransom payments, trust must be established between both victim and attacker.~\cite{Cartwright2019a,Cartwright2019b,Li2020}.



\begin{figure}[h!]
\renewcommand{\figurename}{Algorithm}



%




    \centering
        \includegraphics[scale=.558]{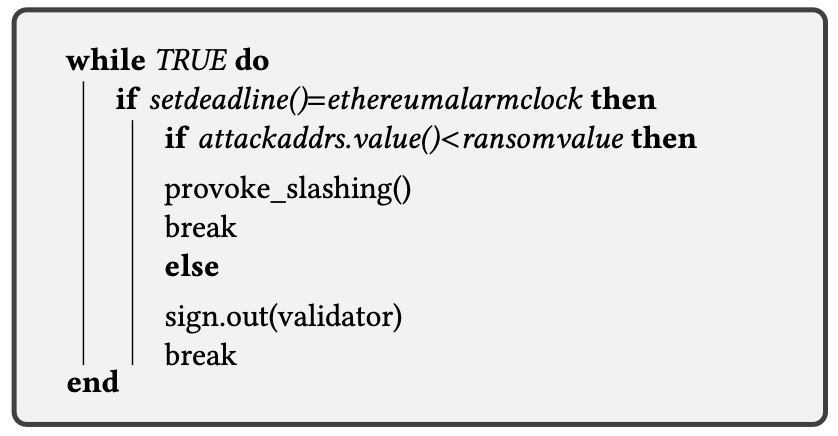}

\caption{Logic implementation of a smart contract by an attacker.} 
\label{algo1}

\end{figure}

Smart contracts offer a way to potentially resolve the need for trust between the victim and attacker \cite{Delgado2020,bhudia2021ransomclave}. To fully resolve the `trust problem', however, a complete contract needs to be devised \cite{Meier2021} and it can be difficult to devise complete contracts that resolve all uncertainty \cite{Baker2005,Hart2017}. We demonstrate that the Ethereum validation process can be exploited by attackers in a way that allows them to write a smart contract, that is complete and incentivizes payment through mechanisms familiar to the victim. The most realistic contract we envisage is illustrated in Algorithm \ref{algo1}. Given smart contracts do not run automatically, an attacker must use an Oracle to trigger the stateful functions to execute the contract's code \cite{Ethereum48:online}.

If a ransom is not paid, then the criminal will provoke slashing. If a ransom is deposited, then the validators will be voluntarily (without penalty) signed to exit. Crucially, this exit is publicly observable and restores control of the eth balance to the victim upon finalisation. The victim would be willing to pay a ransom if that guarantees the exit of their validator(s), and if the cost of ransom is lower than the cost (direct and opportunity) of being slashed. This, in turn, means criminals can exploit this desire to minimise losses, for financial gain. 

\makeatletter

While the contract set out in Algorithm \ref{algo1} may seem simple there are a number of important nuances that would need to be considered in writing and designing a fully watertight contract. For instance, a validator can be slashed in Ethereum while in the exit queue. The victim should, thus, refuse to sign a smart contract that does not guarantee exit from the chain is fully resolved, with no slashing, before the ransom is transferred to the attacker. Such nuances mean that a malicious attacker could potentially manipulate and misinform the victim by, for instance, getting them to sign a smart contract where the ransom is transferred to the attacker once the validators are signed to exit. In this case, the attacker could extract a ransom and subsequently slash the victim's validators, imposing large costs on the victim. Our results help inform on the potential pitfalls faced by the victim so that they can avoid being manipulated by the attacker in this and similar ways. 

One should also note that the smart contract does not need to be created or maintained by either party (attacker or victim) involved in the extortion attempt. In fact, it may even be the case that a third party, as an extension of Ransomware as a Service (RaaS), may operate the contract as a form of neutral third party mediation service (or escrow). 

\subsection{Ethical considerations}

Our paper and analysis follow a process of responsible disclosure. In particular, we highlight a form of attack that is viable and likely, while omitting practical details that would assist criminals. We have disclosed our findings to the DevOps team at the Ethereum Foundation. There are non-trivial integration tasks that would be required for criminals to obtain signing keys, and implement a fully-fledged smart contract to extort victims as envisaged from our model. Our paper does, however, offer practical guidance for potential victims. We believe this form of extortion can be highly damaging for victims and so it is vital to be `ahead of the game' and pre-empt attack. In \S\ref{mitigations} we discuss possible mitigation measures in more detail.

\section{Model of Ethereum validation}
\label{section2}

In this section, we propose a simplified model of the Ethereum (eth) validation process. The model is designed to capture the most salient points relevant to modelling a ransom attack. The notation used in the model is summarised in Table  \ref{tab:notation}.

\begin{table*}[h!]
\caption{Summary of notation used in the model.}
\label{tab:notation}
\renewcommand{\arraystretch}{1}
\begin{tabular}{|l|p{153mm}|}
\hline
\textbf{Variable} &  \textbf{Description}\\ [0.9ex]
\hline
\hline
$t$ &  Time in epochs\\
$\tau$ & Time of special penatly\\
$V$ &  Set of validators (128 nodes in a slot)\\
$V(t)$ & Set of validators active at time $t$\\
$N$ & Set of stakers\\
$\rho$ & Mapping from stakers into a set of validators\\
$y_j(t)$ & Balance of validator $j$ at time $t$\\
$\overline{y}_j(t)$ & Effective balance of validator $j$ at time $t$\\
$Y(t)$ & Total effective balance in period $t$\\
$s_i(t)$ & Stake of $i$ at time $t$\\
$L_j(t)$ & Relative stake of validator $j$ at time $t$\\
$b_j(t)$ & Base reward of validator $j$ at time $t$\\
$\alpha$ & Scale factor reward for matching source, target and head\\
$\beta$ & Scale factor penalty for not matching source, target and head\\
$\gamma$ & Scale factor for offline penalty \\
$l_j(t)$ & Dummy variable that takes value 1 if $j$ performs a slashable offence in period $t$.\\
$L(t)$ & Set of recently slashed validators from period $t$\\
$G(t)$ & Recently slashed balances \\
$Z$ & Delay before slashed validators can exit\\
$\delta$ & Scale factor for per-epoch penalty for slashing \\
$\Delta$ & Scale factor for a special penalty for slashing \\
$P_j(t)$ & Special penalty for validator $j$ if slashed in period $t$\\
$\omega$ & Whistleblower reward \\
$C$ & Set of compromised validators \\
$R$ & Ransom demand \\
$t_r$ & Ransom cut-off time \\
$t_e$ & Time for exit from chain \\
$f$ & Transaction cost for the victim of raising the ransom\\
$\zeta$ & Cost to an attacker for performing actions that force slashing\\
$W$ & The foregone gains from a victim having to wait for validators to exit by inactivity\\
$\overline{R}_E$ & Maximum ransom victim is willing to pay in Ransom and Exit game\\
$\overline{R}_S$ & Maximum ransom victim is willing to pay in the Slash and Ransom game\\
$\overline{R}_P$ & Maximum ransom victim is willing to pay in the Pay or Slash game\\
$H(t_r)$ & Total penalty if slashed at time $t_r$\\
$q$ & Probability attacker has compromised additional validators\\
$Q$ & Size of the total balance of additional compromised validators\\ 
 
\hline
\end{tabular}
\end{table*}


In eth blocks are added to the Beacon Chain through a system of proposal and attestation by validators. Each validator is required to deposit an initial stake of 32ETH.\footnote{A validator can stake more but the effective balance is capped at 32ETH. Encouraging those with $>32$ETH to fund further validators to maximise returns.} Validators are then rewarded or punished, as we shall discuss below, depending on their actions. Ethereum currently has over 500,000 validators, the aim being to increase this number steadily \cite{StakingETH23:online}. This level of decentralisation is managed by groups of at least 128 validators called committees. Validators are randomly assigned to committees with the task of evaluating what is and isn't part of the beacon and shard chains. Committee votes are aggregated into an attestation. This process of aggregation means the validity of the Beacon Chain can be considered without continuously evaluating the votes of many validators, increasing efficiency.

The Beacon Chain is the core of Ethereum in that it manages validators, their duties and the coordination of shards.\footnote{A key feature of Ethereum is that it splits the chain into shards to limit the amount of data that any node needs.} Validator clients handle the logic of a validator. They communicate with the beacon node to understand the state of the Beacon Chain, attest to and propose blocks when appropriate, and ask the beacon node to send the information. To operate on the Beacon Chain, a validator uses a private \emph{signing key} or validator key. Crucially, anyone with the signing key can operate as if they were the validator. 

Alongside the signing key, a validator will have a separate private \emph{withdrawal key}. This key is required to withdraw funds from Ethereum. The separation between the signing key and withdrawal key is important because it means that an attacker who accesses a validator's signing key cannot directly access the validator's stake. Moreover, the withdrawal key, unlike the signing key, does not need to be in active use online and so can be kept more secure in offline storage. In this paper, we model the most likely threat scenario in which an attacker has accessed the validator's signing key but not the withdrawal key.\footnote{The signing key can be recovered from the withdrawal key and so the validator, even if the signing key is compromised during an attack, can recover and still operate.} 

Time on Ethereum is subdivided into \emph{slots} which last 12 seconds each. A slot is the time period in which a new block is expected to be added to the chain. An \emph{epoch} consists of 32 slots and hence lasts 6.4 minutes. This is the unit of time in which each validator is called to make exactly one attestation \cite{Upgradin33:online}. It is also the unit of time within which rewards and penalties are issued. We measure time in our model with epochs, and denote it by $t=0,1,2,...,T$. We define a set of validators as a group or a committee of at least 128 validators (randomly selected in each slot) from which one validator is selected as a block proposer. The remaining 127 validators vote on the proposal and attest to the transactions. Once a majority consensus is reached, the block is added to the blockchain and the block proposer receives a reward, i.e. variable amount of ETH based on a formulaic calculation \cite{ethhub:online}.

We take as given an initial set of $v$ \emph{validators} $V=\{1,...,v\}$ and a set of $n$ \emph{stakers} $N=\{1,...,n\}$. A single staker can own multiple validators. Let $\rho:N\rightarrow V$ denote a correspondence mapping from the set of stakers into the set of validators. An interpretation is that $\rho(i)$ is the subset of validators owned by staker $i$. We distinguish an \emph{active subset of validators}. With a slight abuse of notation, let $V(t)\subseteq V$ denote the subset of validators that are active at time $t$. For each validator $j\in V$ we identify a \emph{balance} $y_j(t)$ and \emph{effective balance}  $\overline{y}_j(t)$ in period $t$. The nuance of the effective balance is not crucial here but we note that the effective balance is capped at 32ETH and is always an integer amount slightly less than the balance.\footnote{To limit fluctuations in effective balance (which are computationally costly) a number of rules are applied. The effective balance will only increase if the validator's balance is 1.25ETH higher than the current effective balance, e.g. if the effective balance is 27ETH, it will only become 28ETH once the balance goes above 28.25ETH. Effective balance only decreases if the validator's balance falls 0.25ETH below the current effective balance, e.g. if the effective balance is 27ETH, then it will only become 26ETH when the balance drops below 25.75ETH.} We denote by $s_i(t)=\sum_{j\in \rho(i)}y_j(t)$ the stake of staker $i$ at time $t$. Let $Y(t)=\sum_{j\in V(t)}\overline{y}_j(t)$ be the total effective balance in period $t$.

In every epoch, each validator is asked to perform either an attestation or a block proposal. Rewards are given for actions that help the network reach consensus. Minor penalties are given for actions that hinder consensus. Slashing only occurs as a response to malicious actions. 
The payoffs of validator $j\in N$ in period $t$ are calculated relative to a base reward~\cite{Beck2020}. The base reward of validator $j$ in period $t$ is given by:
\begin{equation}
    b_j(t) = \frac{\text{BaseRewardFactor}}{\text{BaseRewardsPerEpoch}\sqrt{Y(t)}}=\overline{y}_j(t)\frac{64}{4\sqrt{Y(t)}}
\end{equation}
The base reward is given in Gwei, where $1$ETH is equal to $10^{9}$ Gwei. Rewards and penalties are subtle but not crucial to our model and so we model them approximately. In each epoch, we assume a validator is asked to perform a task. If they perform the task correctly, they receive a payoff of $\alpha b_j(t)$ where $\alpha$ is a scale factor, currently approximately 3. If they fail to match, they receive a penalty of $\beta b_j(t)$ where $\beta$ is a scale factor, currently equal to 3. If validator $j$ fails to perform duties because of being offline or similar, then $j$ receives a penalty of $\gamma b_j(t)$ where $\gamma$ is a scale factor, currently equal to $1$.
\subsection{Slashing}
Of critical importance is the role of slashing. Let $l_j(t)$ be a dummy variable that takes value $1$ if validator $j$ performs a slashable offence in period $t$ and $0$ otherwise. If $l_j(t)=1$ then $j\not \in V(t')$ for all $t'>t$. In other words, a validator who is slashed exits the active set. If $l_j(t)=1$ then $j$ incurs a penalty in epoch $t$, and each of the next $Z=2^{13}=8,192$ epochs (approximately 36 days). The penalty in epoch $t$ is $\overline{y}_j(t)/32$. The penalty in subsequent epochs is $\delta b_j(t)$, where $\delta$ is a parameter currently equal to 3. Validator $j$ also incurs a `special penalty' in period $\tau = t+Z/2$. The size of the special penalty depends on the number of other validators that have been recently slashed. Specifically, let $L(\tau)$ denote the set of validators that have been slashed in periods $\tau-Z+1,...,\tau$. That is, $j\in L(\tau)$ if and only if $l_j(t)=1$ for some $t\in\{\tau-Z+1,...,\tau\}$. Let $G(t)=\sum_{j\in L(t)}\overline y_j(t)$ be recently slashed balances. The special penalty is:
\begin{equation}
    P_j(t) =\overline{y}(\tau) min\left[ \frac{ \Delta G(\tau)}{Y(\tau)},1\right]
\end{equation}
where $\Delta$ is a scale factor currently equal to 3. The important observation here is that the special penalty is based on effective balance (not base reward) and can potentially be large or small depending on the extent of slashing. In particular, if $G(\tau)\geq Y(t)/\Delta$, meaning a third of validators were slashed, then the special penalty is equal to the effective balance. In other words \emph{the validator can lose all their stake}. Conversely, if $G(\tau)$ is small, meaning few validators were slashed, the special penalty can become effectively zero.
 
A slashable offence needs to be caught and reported. This occurs when a whistleblowing validator creates and propagates a message containing the slashable offence. Both the whistleblower and relevant proposer receive a small reward for reporting the offence. We denote by $\omega$ the whistleblower reward. The proposer reward is currently set at $\omega/7$.  The whistleblower reward is offered on a first-come-first-served basis of reporting the offence. To the best of our knowledge, there is no way that an attacker who forces slashing upon a validator could guarantee being the first validator (using another validator) to whistleblow that slashable offence.   

The final preliminary we need to consider is that of validator exit. There are three ways in which a validator can exit the active subset of validators. If the validator balance falls below 16ETH (through inactivity or non-slashable punishment), then they are forced to exit the validator set (without incurring any additional penalties). This is likely to be very rare. If the validator performs a slashable offence then, as discussed above, it exits the active set (and incurs potentially significant penalties). Slashed validators are forced to exit the network after a period of approximately 36 days. Finally, and crucially the validator can signal an intent to stop validating by signing a voluntary exit message. Prior to the Ethereum upgrade being rolled out, there were limits on the withdrawal of funds. However, as part of the Shanghai upgrade \cite{Ethereum46:online} released earlier this year, funds can now be withdrawn once a validator has exited, so we assume that to be the case in our model. 


Voluntary exit is irreversible. A validator who exits would need to withdraw the stake and restake in a new validator to resume operating. Unslashed validators need to wait for at least $2^8$ epochs (approximately 27 hours) to become withdrawable. This wait is necessary to check if the validator should be slashed and provide time for appropriate rewards and proof of custody challenges. The wait, however, could be longer. In particular, in order to guarantee continuity of the Beacon Chain the exit of validators is controlled to avoid too many exiting at the same time. If a large number of validators are in the exit queue, it can take weeks or months to exit. 

It is fundamental to recognise that an attacker who has the signing key of a compromised validator can impose slashing penalties through misbehaviour. For example, an attacker could run a compromised signing key in another validator client process simultaneously, which would incur slashing. There is nothing the victim can do to stop this happening. Moreover, during the time between signing to exit and actually exiting, the validator can still be subject to slashing. An attacker can credibly threaten a victim that they will force slashing upon their validators if the victim signs to exit the validators. Indeed, the attacker could set up an automatic `trip-switch' to force slashing upon a victim who signs to exit from the chain. Access to the signing key, thus, affords the attacker significant control over the victim. 





\section{A criminal motivation for extortion}


Our model takes as given that the attacker has obtained a signing key (but not the withdrawal key). Recall that the signing key is in constant use online while the withdrawal key can be kept secure. We, thus, consider this the most plausible scenario. There are a variety of standard threats and breaches that could result in criminal access to the validator signing key. For instance, insider threat in which an employee of a staking pool acts maliciously. Indeed the relative infancy of the PoS mechanism, and consequent inexperience of stakers and under-developed protocols of staking pools, would suggest security may be relatively lax providing `low hanging fruit' for criminals.

An attacker holding a validator’s leaked key could mount long range attacks \cite{Deirmentzoglou2019}. This, however, has two major disadvantages: (a) It is highly unlikely to succeed because the attacker would need to control an implausibly large number of validators to have any realistic chance of compromising the blockchain. Ethereum, and other proof-of-stake mechanisms are set up to defend against such long range attacks. (b) Criminals typically like to ‘fly under the radar’ and profit without drawing undue attention. Manipulating the blockchain, while it could be highly profitable if pulled off, would draw considerable attention and likely, fundamentally undermine the currency. Undermining a currency is unlikely to be directly profitable to a criminal gang. 

We will show that extorting validators or staking pools can be highly profitable without attracting attention. Morevoer, it is possible to profitably extort even a single validator (where there would be no chance of a viable long range attack). Criminals could, thus, target validators and staking pools, breach those with the most lax security, and use profitable extortion techniques. In the next section we outline one viable method of extortion.

\section{Pay and Exit Strategy}
\label{section3}
In this section we set out a potential strategy a savvy attacker could use to ransom a validator. We also show how the payment of a ransom could be fully incentivized through a smart contract. We consider the case where an attacker has fully compromised the signing keys of a set of validators at time $t=0$ (the withdrawal key is not compromised). Let $C \subseteq V$ denote the set of validators that have been compromised. By that we mean that: the attacker has the signing keys and so can perform actions that would be expected to result in the validators being slashed.  It is simplest to consider the case in which the set of compromised validators $C$ is owned by a single staker $i$. In other words $C\subseteq \rho (i)$. We explore the extent to which the attacker can ransom the staker. We take as given that an insider is acting maliciously and/or a cyber breach has led to the signing key becoming known to external actors.

\tikzset{
    solid node/.style={circle,draw,inner sep=1.5,fill=black},
    hollow node/.style={circle,draw,inner sep=1.5}
}

To formally model the interaction between attacker and victim we introduce the following dynamic game, which we refer to as the \emph{Pay and Exit game}. It is summarized in Figure \ref{fig:exitinactivity} and can be explained as follows:

\setcounter{figure}{0}
\begin{figure*}[h!]
\renewcommand{\figurename}{Figure}

    \centering
        \resizebox{\textwidth}{!}{%
\begin{tikzpicture}
    [grow=right,
    sloped
    ]
    \tikzstyle{end} = [shape=rectangle, rounded corners,
    draw, align=center,
    top color=white, bottom color=blue!20]
    \tikzstyle{level 1}=[level distance=30mm,sibling distance=50mm]
    \tikzstyle{level 2}=[level distance=35mm,sibling distance=45mm]
    \tikzstyle{level 3}=[level distance=30mm,sibling distance=25mm]
  \node(0)[end]{$\text{Attacker}$}
   child{node(1)[end]{$\text{Victim}$}
       child{node(2)[end]{$\text{Attacker}$}
          child{node[hollow node,label=right:\small{$\bigg[-\zeta,\sum_{j\in C}y_j(t_r)-H(t_r)\bigg]$}]{}
             edge from parent node[below]{$\small\text{Force slashing}$}}
          child{node[hollow node,label=right:\small{$\left[0,\sum_{j\in C} y_j(t_r)\right]$}]{}
             edge from parent node[above]{$\small\text{Not slash}$}}
          edge from parent node[below]{$\small\text{Not deposit \& Exit}$}
       }
       child{node(3)[end]{$\text{Attacker}$}
          child{node[hollow node,label=right:\small{$\bigg[-\zeta,\sum_{j\in C}y_j(t_r)-H(t_r)-f\bigg]$}]{}  
            edge from parent node[below]{$\small\text{Force slashing}$}}
          child{node[hollow node,label=right:\small{$\left[R,\sum_{j\in C} y_j(t_e)-f-R\right]$}]{}
            edge from parent node[above]{$\small\text{Not slash}$}}
          edge from parent node[above]{$\small\text{Deposit R \& Exit}$}
       }
       edge from parent node(2)[above]{$\small\text{sets R}$} node(2)[below]{\& $t_r$}
   };
\end{tikzpicture}%
        }
\caption{A reduced form game tree for the Pay and Exit game.} 
\label{fig:exitinactivity}
\end{figure*}
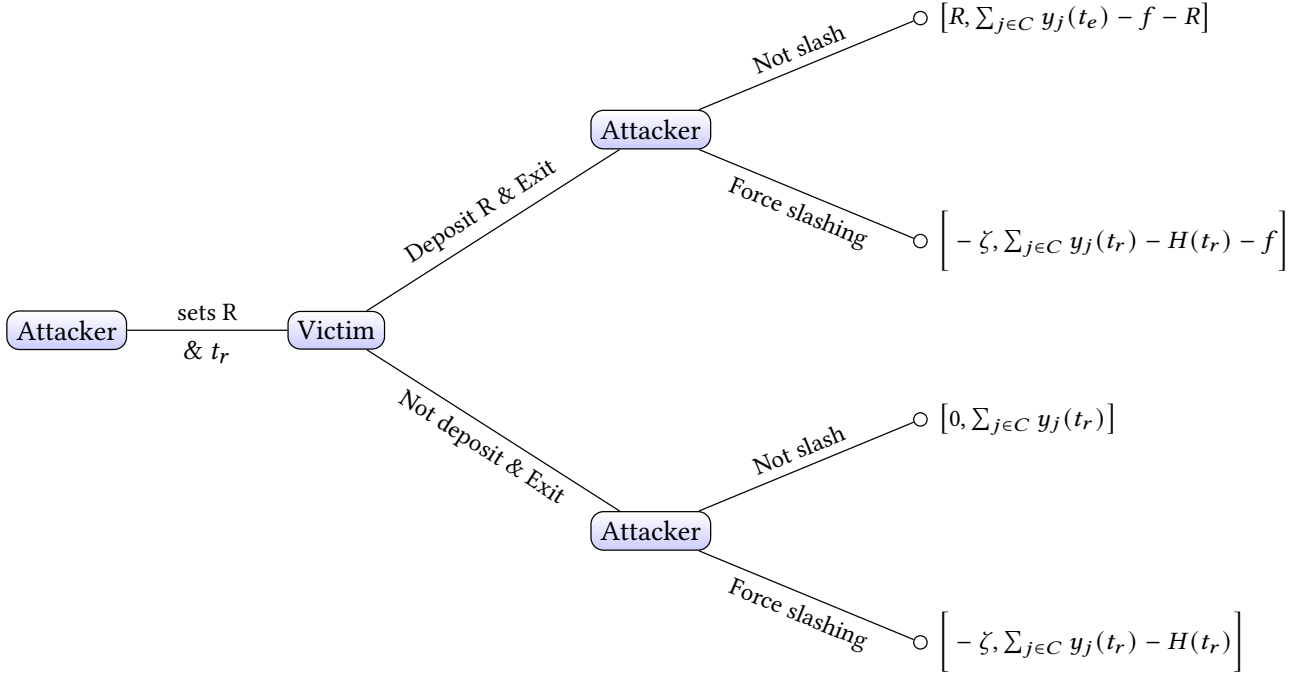

\begin{itemize}

    \item At time $t=0$ the attacker sets a ransom demand $R$ and a time $t_r$ by which the ransom must be paid. He also initiates a self-enforced smart contract.

    \item At time $t=1,2,...,t_r$ the victim decides whether to pay or not. If she decides to pay, then she deposits the required ransom demand $R$ and signs to exit the validators before time $t_r$. She incurs transaction costs $f\geq 0$ which reflect the potential fees from raising and transferring funds.\footnote{These costs could depend on time, e.g. it is more expensive to raise funds quickly, but we abstract from this here.} 
  
    \item At time $t_r$ the attacker chooses whether to force slashing upon the validators or not. If the victim deposits the ransom and the attacker does not slash then the validators exit at time $t_e$. If the attacker forces slashing, he incurs a cost of $\zeta$ which is the cost of operating the validator net of any expected whistleblowing reward. We allow that $\zeta<0$ meaning the expected whistleblowing reward fully offsets the costs of the attack (smart contract/oracle execution) and the criminal materially gains from slashing.  
    
    \item The smart contract dictates that if the ransom is deposited by the victim, the validator(s) are signed to exit, and the validator(s) exit without being slashed then the attacker receives the ransom. If the validator(s) are not signed to exit or slashed (either before or after exit), then the deposit is returned to the victim. This smart contract can be enforced because signing to exit and slashing are observable on the Beacon Chain by the Oracle, which can then trigger appropriate smart contract code.
\end{itemize}

We omit other actions the victim and attacker could perform, such as the victim signing to exit without paying the ransom, because they complicate analysis without affecting our conclusions. In the case of the victim signing to exit, this will be observed by the attacker, who can, within approximately 27 hours, provoke slashing to punish the victim. This does not functionally alter the outcome of the attack, as it is non-compliant behaviour that the attacker is motivated to punish. 

Critical to determining the payoff to each outcome is the total penalty from slashing given by:
\begin{equation}
    H(t_r)=\sum_{j\in C} \left(\frac{\overline{y}_j(t_r)}{32}+\delta\sum_{t=t_r }^{t_r+Z-1} b_j(t)+P_j(t_r)\right)
\end{equation}

Various components of this penalty cannot be known for certain at time $t_r$ or before. For instance, $P_j(t_r)$ depends on the proportion of balances that will be slashed in the period after $t_r$. The attacker and victim can, though, approximate payoffs with a reasonably high level of accuracy based on the history of the Chain. Payoffs can be summarized as follows:

\begin{itemize}
    \item If the victim deposits the ransom and the attacker does not slash then the attacker has payoff: 
    \begin{equation}
        \Pi(\text{Deposit},\text{NotSlash})=R
    \end{equation}  given by the ransom. The victim has payoff: \begin{equation}
        U(\text{Deposit},\text{NotSlash})=\sum_{j\in C} y_j(t_e)-f-R
    \end{equation} given by the balance withdrawn at the time of exit $t_e$ minus the transaction cost of the deposit and the ransom. The validator balance at time $t_e$, i.e. $y_j(t_e)$, can be approximated using $y_j(t_e)=y_j(0)$.
    \item If the victim deposits the ransom and the attacker forces slashing, then the attacker has payoff: \begin{equation}
        \Pi(\text{Deposit},\text{Slash})=-\zeta
    \end{equation} where $\zeta$ is the cost the attacker incurs from forcing slashing. The victim has payoff: 
    \begin{equation}
        U(\text{Deposit},\text{Slash})=\sum_{j\in C} y_j(t_r)-H(t_r)-f
    \end{equation}  to reflect the loss of balance from slashing. The smart contract means the ransom is returned to the victim.
    \item If the victim does not deposit the ransom and the attacker does not slash in period $t\leq t_r$ then we assume the attacker's and victim's payoff are:\footnote{In this scenario the game continues because the validators are still not exited. For now we abstract from this and assume that the fact the attacker did not force slashing makes any future threat non-credible.} 
    \begin{equation}
        \Pi(\text{NotDeposit},\text{NotSlash})=0 
    \end{equation} 
    and
    \begin{equation}
        U(\text{NotDeposit},\text{Sign})=\sum_{j\in C} y_j(t_r)
    \end{equation}
    \item If the victim does not deposit the ransom and the attacker forces slashing then the attacker has payoff: \begin{equation}
    \Pi(\text{NotDeposit},\text{Slash})=-\zeta
    \end{equation} 
    The victim has payoff: \begin{equation}
        U(\text{NotDeposit},\text{Slash})=\sum_{j\in C} y_j(t_r)-H(t_r)
    \end{equation}
\end{itemize}

As a solution concept we look for a sub-game perfect Nash equilibrium~\cite{Cartwright2019b}. This requires all actions on the equilibrium path to be individually rational. We can solve for the sub-game perfect Nash equilibrium using backward induction. If $R>-\zeta$, then it is optimal for the attacker to sign to exit if the ransom is deposited. If the ransom is not deposited, then it is optimal for the attacker to force slashing if $-\zeta>0$. The victim can, thus, predict that their payoff if they deposit is $U(\text{Deposit},\text{NotSlash})$ and their payoff if they do not deposit is $U(\text{NotDeposit},\text{Slash})$. It is, therefore, optimal for the victim to deposit the ransom if:
\begin{equation}
    \sum_{j\in C} y_j(t_e)-f-R>\sum_{j\in C} y_j(t_r)-H(t_r)
\end{equation}
For simplicity we assume $t_e\approx t_r$. This, yields a maximum ransom the victim is willing to pay of:
\begin{equation}
    R < H(t_r)-f=\overline{R}_E
    \label{eq:exitransom}
\end{equation}
If $\overline{R}_E>-\zeta$ then the unique subgame perfect equilibrium of the game is for the attacker to set ransom $R=\overline{R}_E$, the victim to deposit the ransom and the attacker to sign the validator(s) to exit. Thus, it can be fully incentive compatible for the victim to pay the ransom with the smart contract giving confidence that payment of the ransom will result in a positive gain. In a standard ransomware attack it is unlikely the attacker can accurately predict the willingness to pay off the victim~\cite{Hernandez2020}. By contrast, in this setting, the willingness to pay can be predicted with a much higher level of accuracy.

We highlight that the preceding equilibrium is based on the assumption $-\zeta>0$, meaning that if the ransom is not paid it will be optimal for the attacker to force slashing on the validators. If it is costly, however small, for the attacker to go through the process of forcing the validators to be slashed then there are incentives for the attacker to do nothing (incurring no cost). The threat of slashing may be non-credible under some circumstances. The potential non-credibility of threats in `kidnapping' scenarios is well documented~\cite{Cartwright2019b}. In this instance, credibility is maintained by the attacker providing a cryptographic proof of ownership for the validator(s) staking key. The attacker signs a mutually known string with the private staking key, and allows the victim to decrypt it using their public staking key. Obtaining a matching copy of the known string, proves ownership of the private staking key. The attacker is also motivated to follow through on their threat of slashing, as they may receive a whistleblower reward. This motivates punishment of non-compliant victims, as this may offset the direct cost of the attack (smart contract execution) as well as the maintenance of global credibility through punishing misbehaviour. Given that the costs incurred from forcing a slash are likely to be small, any gains would likely overcome the costs. This would imply $\zeta<0$ and, thus, slashing is a credible threat.


The Pay and Exit game illustrates that an attacker can credibly threaten a compromised validator in a way that incentivises the victim to pay a ransom. Crucially, this strategy is fully incentive compatible and can be written into a smart contract. The smart contract uses the fact that exit without slashing is fully observable and exit is irreversible. Thus, the victim can know that payment of the ransom will prevent losses in excess of the ransom. In practice the optimal ransom will differ from $\overline{R}_E$ because of nuances in payoffs not modelled in the Pay and Exit game, as well as the need to predict variables such as $t_e$ and $H(t_r)$. But these are minor issues that do not challenge the notion that the Pay and Exit strategy is effective for the attacker. We also consider the Pay and Exit strategy to be an intuitive and simple approach for the attacker. 

\section{Pay or Slash Strategy}
\label{section6}

In this section we briefly consider a strategy that may seem a natural starting point for criminals but is unlikely to be optimal. In this instance, the attacker claims that if a ransom is paid by the victim then the attacker will not act maliciously against the victim. If a ransom is not paid the attacker will act so as to force slashing upon the compromised validators. This approach has the advantage that the victim would not have to exit the validators, as in the Pay and Exit strategy. This is advantageous given that exit involves some costs and delays the victim wants to avoid. The key question is whether the claim by the attacker that they would not act maliciously, beyond the remit of the smart contract, is credible.

In order to obtain a complete contract that could be enforceable in a smart contract, there must be some observable measure of not acting maliciously. The only natural candidate is that the validators are not slashed and operate as normal. To capture this, we envisage a smart contract of the form: \emph{A ransom will be transferred to the attacker if the validators are not slashed in the next X epochs}. We call this the Pay or Slash strategy. Crucially, a time limit of X epochs is necessary in order to make this contract workable. 
While the time limit is necessary, it creates an obvious problem. On epoch $X +1$ there is nothing to stop the attacker from renewing a threat to slash unless a second ransom is paid. We must model an extended game in which the threat to slash is repeated until, ultimately, the compromised validators exit, either voluntarily without slashing or with slashing or because Ethereum stops operating. 

A victim may be willing to pay a ransom to delay malicious actions. We can put a cap on the cumulative amount of ransom the victim would be willing to pay in any sub-game perfect Nash equilibrium. Specifically, suppose the victim does not deposit the ransom by the initial predefined time limit of $t_r$ and incurs slashing as a consequence. This would result in a predicted loss from the slashing penalty of $H(t_r)$. The cumulative ransom that the victim would provision to pay in any equilibrium strategy cannot, therefore, exceed $H(t_r)$. Otherwise, the victim would have a higher expected payoff by signing to exit the validators, in the hope they avoid slashing. Once we include the costs of depositing the ransom we see that the cumulative ransom cannot exceed that from the Pay and Exit strategy, $\overline{R}_E$.

If in equilibrium, the cumulative ransom paid over successive iterations of the ransom demand cannot exceed $\overline{R}_E$ we would expect the ransom paid in each iteration to be well below $\overline{R}_E$. The more iterations of the game, the lower the ransom one would expect in any one iteration along the equilibrium path. There is also the issue of unchangeable cryptographic keys remaining in the possession of the attacker. As the attacker can never provide compelling proof that they have destroyed the private key(s) in their possession, the victim must trust that they will honour the X epoch limit between attacks. They must also trust that the attacker will not disclose, sell, or otherwise share this key. This is an unsustainable, one-sided trust relationship. Clearly, it is preferable, from the perspective of both the victim and attacker, to employ a Pay and Exit strategy. This resolves the problem more quickly and means both parties have payoffs at least as large as with the Pay or Slash strategy. It also implicitly proves that the attacker cannot posses the new private key. The Pay or Slash strategy is therefore inferior to the Pay and Exit strategy, especially from the victim's perspective, who onboards additional risk and cost for a questionable degree of convenience.

That the Pay or Slash strategy is not optimal does not mean it would not be tried by attackers. Indeed, once we account for `human factors’ the Pay or Slash strategy may be advantageous for attackers. This is because victims may `mistakenly’ pay a ransom near $\overline{R}_E$ in the first iteration in the `hope’ this will make the problem go away. Once that first ransom is paid the attackers can renew their threat and secure subsequent ransoms. Our analysis, therefore, offers some direct advice that can be given to validator owners to avoid such eventualities. Specifically, any victim considering paying a ransom to the attackers should request a Pay and Exit strategy. This secures the victims and allows them to move on from the attack. While it involves payment of a ransom and so is not ideal, it is better than alternative strategies that would also result in paying a ransom.


\section{Discussion}
\label{discussionsection}
\label{section7}
In this paper we have shown proof of concept that an attacker could exploit the validation process of Ethereum for financial gain. There are various methods an attacked could use to compromise the signing key of a validator if the staker has lax security. An attacker who has compromised the signing key of a validator could hold the staker to ransom. We propose a simple and effective strategy the criminal could use, called the pay and exit strategy. The attacker threatens to force slashing upon the validators unless a ransom is paid. If the ransom is paid then the compromised validators will be allowed to voluntarily sign to exit. Crucially, exit is irreversible and so returns control of the stake to the victim. The ransom payment can be made conditional, in a self-enforcing smart contract, on the validators exiting without slashing.   
    
A rational victim would be willing to pay a ransom to stop the threat being carried out. Moreover, we have shown that a smart contract could be enacted between the victim and attacker that would make payment of the ransom incentive compatible for the victim. Of the two strategies we have considered Pay or Slash is the least effective (from an equilibrium perspective) because it results in a lower ransom payment and a more prolonged process. Arguably, it is the most likely to be first seen in the wild given that it is an ‘obvious’ attack strategy. Moreover, `naive' victims may pay irrationally large ransoms in a process of `learning by error'. Pay and Exit is intuitive and preferable for both attacker and victim in that it leads to a straightforward resolution of the problem without any slashing. Note, however, that here there is still important nuance that an attacker could exploit, for instance the potential for validators to be slashed between signing to exit and actually exiting the Chain.  

We make no claims that the two strategies outlined above are the only ones that could be used by attackers. Our objective was to highlight that a threat exists and that it could take various forms that would result in a credible ransom that the victim is motivated to pay. We have focused on games and smart contracts that are built around easily observable and transparent outcomes - e.g. a validator is signed to exit or the recently slashed balances are below a pre-set threshold. It may be possible, to design smart contracts around `less transparent' actions, particularly with the help of oracles such as Chainlink \cite{chain46:online}. Relevant here is the potential for an attacker to trade payment of the ransom for handing back control of the signing keys. The problem with this approach (familiar from the ransomware literature \cite{Delgado2020}) is how to make sure the correct signing keys are returned and that the attacker cannot re-corrupt the signing keys or otherwise retain possession of those keys. An attacker signing for the validators to exit is one simple way of overcoming this problem, but there may be other potential solutions.

The modelling framework we have assumed is open to many modifications and extensions. We assume that the attacker sets up and initiates a smart contract with a ransom demand $R$. There is, in principle, no reason why the victim could not set up a similar smart contract with a ransom amount of less than $R$. A negotiation and bargaining process is possible in which offers and counter-offers can be made and enforced via a smart contract. The attacker's ability to extract the full amount that the victim is willing to pay is not, therefore, guaranteed. This does not change the main result that rational victims have an incentive to pay a ransom and attackers have an incentive to attack validators. We are merely clarifying that the attackers may not be able to fully exploit the victims' willingness to pay if negotiations ensue.

It is important to explore the extent to which ransom attacks on validators could disrupt Ethereum. For the most part an attack will have a little direct effect on Ethereum as a whole. For instance, if a single validator is compromised then, at most, the validator pays a ransom and exits or does not pay and is slashed. This can be profitable for the attackers and disruptive for the validator but is unlikely to significantly disrupt the Ethereum network given that it has hundreds of thousands (and soon possibly millions) of active validators. One can foresee a potentially  larger threat to Ethereum from extortion and ransom of much larger sets of validators hosted in large providers. This is especially true if validators owned by a single operator are hosted on shared hardware (e.g., many validators on one server blade). Most, if not all, other PoS networks which do not employ delegated staking, would be exposed to a similar threat, even if the smart contract design and the ransom amounts differ.

The direct threat is that an attack on a staker with a significant proportion of validators could be highly disruptive. This is another reason to further encourage decentralisation and client/software diversity. We remind that the penalty from slashing is more than 3 times the ratio of recently slashed balance to total balance. So, if say, a staker with a sixth of the validators is compromised the potential impact on Ethereum is huge, both in terms of disruption to activity and potential loss of assets. 
That would clearly have a significant impact on Ethereum's quality of service, reputation, and future. In assessing whether or not this scenario is realistic, we detail in Table \ref{tab:distribution} the current distribution of the larger validators. As shown in Table \ref{tab:distribution}, there is increasing adoption of the new PoS Ethereum network.

\begin{table}[h!]
\begin{center}
\caption{Distribution of Ethereum validators by deposits as of March 2023 \cite{Breakdown92}. }
\label{tab:distribution}
\renewcommand{\arraystretch}{1}
\begin{tabular}{|p{24mm}|p{23mm}|p{23mm}|}
\hline
\renewcommand{\arraystretch}{10}
\textbf{No. of validators} & \textbf{No. of stakers March 2023} & \textbf{No. of stakers Nov 2021} \\ [0.3ex]
\hline
\hline
1 & 83,365 & 40,144 \\
2-5 & 4,393 & 2,341 \\
6-10 & 882 & 442 \\
11-50 & 2,179 & 1,376 \\
51-100 & 231 & 92  \\
101-500 & 268 & 103 \\
501-1000 & 47 & 28  \\
$>$1000 approx & 60 & 42 \\
\hline
\end{tabular}
\end{center}
\end{table}

Though a significant proportion of stakers (representing unique wallets) have funds staked against a single validator, a small number of stakers (approximately 0.01\% of the total population) staked funds in over 1,000. This indicates a variety of strategies, both for risk mitigation and the normalisation of gains. A wallet with only a single stake could be the sole (32ETH) staker in a given validator, or they may have a lower net worth and instead be attempting to gain interest on their <32ETH holdings. Stakers putting funds into a great many validators may be distributing their risk and/or attempting to ensure returns as validator selection is effectively random, so a more diffuse stake across all validators is a sound strategy if one seeks consistency in returns. One can theoretically profile all such stakers and the amounts they have staked against validators, but as attackers (in most cases) lack personal or forensic information that would tie multiple wallets to individual off-chain identities, Table ~\ref{tab:distribution} demonstrates that focusing on individual stakers, as opposed to pool operators, is inefficient. As all validators have 32ETH, and we are unable to extract funds directly in our model attack, the game presented in this paper will always be played against the owner of a validator, not the individual stakers. The diversity of investment strategies and variable number of stakers associated with a validator have no impact on the game presented in this work - pool operators may interact with their stakers or associates in various ways to decide whether they can or will pay, but this is of no concern to the attacker, whose goals are best served by exploiting any general features of all Ethereum validators (dominant hosting strategies and uniform total funds) assuming no prior knowledge of the target validator(s). 



Ransom attacks may also disrupt Ethereum through the indirect impact on stakers' willingness to act as validators. A proof of stake system relies on a large number of independent validators. Hence, anything that discourages people or organisations from acting as validators is a concern. If ransomware attacks become prevalent or validators become easy targets for attackers, then this discourages investment because it provides a clear downside risk. Moreover, there is an inevitable element of competition between cryptocurrencies in trying to attract investment \cite{Gandal2016}. This means that `bad publicity' could have a particularly damaging effect on efforts to attract a wide range of diverse validators. A publicly reported and successful attack on a validator could be the catalyst for a rapid escalation in the frequency, extent, and sophistication of future attacks.

\section{Mitigation}
\label{mitigations}

Having identified the extortion threat to validators we conclude by discussing various mitigation measures that can be employed by stakers and Ethereum to limit the threat of extortion. We summarize these as follows:

\begin{enumerate}[]
\item Security of the signing key is paramount in stopping extortion. Validators and staking pools should, thus, take active measures to secure the signing key. This would include counter-measures for external threat (virus protection etc.) and internal threat (restricted access and employee vetting etc.). While these measures would seem standard, the relative infancy of the sector, and its consequent lack of maturity, suggest that security may be lax. A further consideration is the diversity of individuals who run validators. The broad spectrum of technical ability and understanding represented within the Ethereum staking community leads to an unpredictable and heterogeneous level of protection, on a per validator basis. 
\item Investors should be encouraged to check and verify the security credentials of staking pools. In a competitive market system, the security controls of staking pools will reflect the preferences of investors, and their willingness to pay for security measures. If investors fail to appreciate the importance of security one can obtain a `race to the bottom' in which staking pools implement lax security as a cost-cutting measure \cite{bhudia2022extortion,bhudia2022identifying}. Investors need education on the importance of using secure staking pools and how to appropriately quantify risk. This can put pressure on staking pools to raise the level of security.  
\item Compromised validators can be advised to not pay a ransom under any circumstances where they are not guaranteed to regain full control over their stake. The Pay and Exit strategy, encapsulated in a smart contract, is an effective way for victims to fully resolve the problem (although security issues may remain from the initial breach). This is not to say that we would encourage victims to pay a ransom under a Pay and Exit strategy. We can, however, strongly discourage payment under any alternative circumstances that do not give similar guaranteed return of control.

\item The threat of slashing is most damaging if a staker with a large number of validators is breached. The natural solution is for a staker to partition their validators (such that lateral traversal is not possible) so that a breach can only ever threaten a portion of the validators. Segmentation, thus, limits the potential losses from extortion. It also has the benefit of increasing the per-validator cost of attack, in the pre-extortion phase for the attacker.

\item The `attractiveness' of validator extortion to criminals will depend on how easily they can breach validators. The harder it is to breach validators the less attractive the crime will appear. Validator security, therefore, has a positive externality effect on other validators by discouraging criminal activity. This means it is in the interests of staking pools and investors to encourage good security in others, reinforcing points (1) and (2). Community standards and advice are critical to establishing minimum operational security across the breadth of Ethereum validators. 

\item The amount that criminals can extort depends on the slashing penalty. Lowering the slashing penalty would, thus, lower the potential gains to criminals and reduce the losses to victims. The slashing penalty is needed to disincentivize malicious activity. The malicious activity of small stakers would, though, have no material impact on Ethereum because, realistically, it cannot corrupt the blockchain. By lowering the fixed slashing penalty, while retaining the special penalty, the extortion threat to small stakers would be dramatically reduced, while malicious activity by large stakers is still disincentivised. The parameters of Ethereum could, thus, be adjusted to reduce the threat. The issue with this approach is that it may also impact the real-terms security of the blockchain itself. It is likely that the Ethereum community would be concerned by a reduction in slashing penalties on these grounds. As an item of future work, we intend to study the current security posture of the Ethereum blockchain, and advise whether the current level of slashing can be safely reduced.
\end{enumerate}

\section{Conclusion}
In this paper, we have analysed how an attacker who has gained access to the validator(s) signing key could threaten to perform malicious actions, such as slashing and extract ransom from them. We used game-theoretic modelling to determine the credible threats that an attacker could exploit to perform extortion attacks. In our study we observed that in all of the counter-measures described in \S\ref{mitigations}, there are trade-offs, e.g. partitioning of validators is costly but reduces the losses from attack. Our analysis shows significant downside risks from stakers not performing robust security actions. In the future, could look at the cost-benefit trade-off to identify the optimal security level and associated level of risk. Another possibility would be to investigate the idea of deploying insurance solutions to address attacks like those described here, so that network disruption is minimised in all circumstances and validator losses can be de-risked.

\subsection{Future work}
It may seem that one could simply implement a countermeasure against slashing-as-leverage for extortion, by allowing withdrawal keys to be used to signal that a validator is compromised, and appeal for slashing penalties to not be enforced. However, this fundamentally undermines the security model of the Ethereum blockchain, and would result in all malicious validators making such claims to reduce the cost of misbehaviour (mitigate slashing). As a result, we are investigating new security measures that could be implemented at a protocol level to mitigate extortion attacks through a key rotation mechanism. The results of the analysis reported in this work have been shared with the Ethereum Foundation, who have funded additional work (Grant \#FY22-0720 ‘REVOKE: Consensus-layer mitigations for validator ransomware attacks’) to explore consensus protocol adaptations to mitigate the impact of this novel extortion attack against proof of stake validators.

\begin{acks}
We thank Justin Drake from the Ethereum Foundation for his support and feedback throughout this research. This project was partly supported by Ethereum Foundation Grant \#FY21-0378 ‘Game theoretic modelling of a ransomware attack on validators in Ethereum 2.0’. The research of Alpesh Bhudia is supported by the EPSRC and the UK government as part of the Centre for Doctoral Training in Cyber Security at Royal Holloway, University of London (EP/P009301/1).
\end{acks}

{}











\bibliographystyle{ACM-Reference-Format}
\bibliography{paperarxiv}

\appendix

\end{document}